\documentclass[runningheads]{llncs}
\usepackage[T1]{fontenc}
\usepackage{graphicx}
\usepackage[show]{chato-notes}
\usepackage{booktabs}
\usepackage{xspace}
\usepackage{pifont}
\usepackage{amsmath} 
\usepackage{booktabs}
\usepackage{multirow}
\usepackage{siunitx}
\usepackage{subcaption}
\usepackage{threeparttable}
\usepackage{hyperref}
\newcommand{\na}{\multicolumn{1}{r}{$-$}} 

\usepackage{amssymb}
\newcommand{\msmarco}{\textsc{MsMarco}\xspace}
\newcommand{\msmarcovo}{\textsc{Ms Marco-v1}\xspace}

\newcommand{\lotte}{\textsc{LoTTE}\xspace}

\newcommand{\seismic}{\textsc{Seismic}\xspace}
\newcommand{\kannolo}{\textsc{kANNolo}\xspace}
\newcommand{\faiss}{\textsc{Faiss}\xspace}
\newcommand{\warp}{\textsc{Warp}\xspace}
\newcommand{\emvb}{\textsc{Emvb}\xspace}
\newcommand{\igp}{\textsc{Igp}\xspace}
\newcommand{\splate}{\textsc{Splate}\xspace}
\newcommand{\plaid}{\textsc{Plaid}\xspace}

\newcommand{\splade}{\textsc{Splade}\xspace}

\newcommand{\colbert}{\textsc{ColBERT}\xspace}
\newcommand{\colbertvt}{\textsc{ColBERTv2}\xspace}
\newcommand{\xtr}{\textsc{Xtr}\xspace}

\newcommand{\bmtf}{\textsc{BM25}\xspace}
\newcommand{\lilsrbig}{\textsc{Big}\xspace}
\newcommand{\cocond}{\textsc{CoCondenser}\xspace}
\newcommand{\lilsr}{\textsc{Li-Lsr}\xspace}

\newcommand{\paroff}{\textsc{off}\xspace}

\begin{document}
\title{Multivector Reranking in the Era of Strong First-Stage Retrievers}
%

\author{Silvio Martinico\inst{1,3}\orcidID{0009-0005-7280-6147} \and
Franco Maria Nardini\inst{1}\orcidID{0000-0003-3183-334X} \and
Cosimo Rulli\inst{1}\orcidID{0000-0003-0194-361X} \and
Rossano Venturini\inst{2}\orcidID{0000-0002-9830-3936}}
\authorrunning{S. Martinico et al.}
%
\institute{ISTI-CNR, Pisa, Italy,
\email{\{name.surname\}@isti.cnr.it} \and
University of Pisa, Italy,
\email{rossano.venturini@unipi.it} \and 
University of Pisa, Italy,
\email{silvio.martinico@phd.unipi.it}
}

\maketitle
\begin{abstract}
Learned multivector representations power modern search systems with strong retrieval effectiveness, but their real-world use is limited by the high cost of exhaustive token-level retrieval. Therefore, most systems adopt a \emph{gather-and-refine} strategy, where a lightweight gather phase selects candidates for full scoring. However, this approach requires expensive searches over large token-level indexes and often misses the documents that would rank highest under full similarity.
In this paper, we reproduce several state-of-the-art multivector retrieval methods on two publicly available datasets, providing a clear picture of the current multivector retrieval field and observing the inefficiency of token-level gathering.
Building on top of that, we show that replacing the token-level gather phase with a single-vector document retriever---specifically, a learned sparse retriever (LSR)---produces a smaller and more semantically coherent candidate set. This recasts the gather-and-refine pipeline into the well-established two-stage retrieval architecture. 
As retrieval latency decreases, query encoding with two neural encoders becomes the dominant computational bottleneck. To mitigate this, we integrate recent inference-free LSR methods, demonstrating that they preserve the retrieval effectiveness of the dual-encoder pipeline while substantially reducing query encoding time.
Finally, we investigate multiple reranking configurations that balance efficiency, memory, and effectiveness, and we introduce two optimization techniques that prune low-quality candidates early. Empirical results show that these techniques improve retrieval efficiency by up to 1.8$\times$ with no loss in quality. Overall, our two-stage approach achieves over $24\times$ speedup over the state-of-the-art multivector retrieval systems, while maintaining comparable or superior retrieval quality. 

\end{abstract}
%
%


\section{Introduction}
\label{sec:intro}

Transformer-based neural architectures~\cite{attention} and large pre-trained language models such as BERT~\cite{bert} have reshaped the field of Information Retrieval (IR), thanks to their ability to encode rich semantic representations of text.  
These models have enabled a shift from purely lexical matching to representation-based retrieval, where queries and documents are mapped into vector spaces that capture deep contextual meaning.  
This paradigm has led to a new generation of neural retrievers that substantially improve retrieval effectiveness over traditional term-based models such as BM25~\cite{bm25}.
These models can be broadly categorised into \emph{sparse} and \emph{dense} encoders.  
Sparse encoders, such as \splade~\cite{splade}, produce high-dimensional representations, where only a small fraction of components are non-zero. 
In contrast, dense encoders represent each query or document as a dense vector in a low-dimensional space that captures global semantic similarity. They can be further divided into \emph{single-vector} and \emph{multivector} architectures.  
Single-vector encoders~\cite{dpr,ance,star-adore} map an entire passage to a single dense vector, providing compact document-level semantics but limiting fine-grained query–document interactions. Differently, multivector models such as \colbert~\cite{colbert} encode each query and document as a set of token-level vectors. In this framework, similarity 
between a query and a document 
is computed via late interaction through the max similarity operator, $MaxSim(q, D) = \sum_{i=1}^{n_\mathtt{q}} \max_{j=1,\dots,n_\mathtt{D}} \langle q_i, D_j \rangle$, which lets every query token interact with every document token~\cite{colbert}. These models achieve state-of-the-art effectiveness because they preserve fine-grained token interactions. However, performing retrieval over a full multivector collection is computationally infeasible. For this reason, all scalable multivector systems follow a \emph{gather-and-refine} strategy~\cite{plaid,emvb,igp,warp}: 1) a \emph{gather} phase that selects a relatively small set of candidate documents; 2) a \emph{refine} phase that applies full $MaxSim$ scoring on these candidates to produce the final ranking.

Most existing systems perform the gather step at the \emph{token-embedding} level. As already noted in~\cite{xtr}, this design has two fundamental drawbacks: 1) the index cardinality (the number of vectors to search) grows with the number of tokens, which is typically one or two orders of magnitude larger than the number of documents, leading to significantly higher search-space complexity and slower queries, and 2) token-level similarity alone does not predict document-level relevance well: $MaxSim$ depends on the joint interaction of all query tokens with all document tokens, and the final score depends on many different document tokens. 
The token-based gather step is therefore both \emph{inefficient}, because it must retrieve and evaluate many more candidates, and often \emph{ineffective}, because it fails to identify the documents that would score highest under full $MaxSim$.  
To mitigate this, XTR~\cite{xtr} redefines the training objective to make token retrieval itself more predictive of document-level relevance, partially closing the training–inference gap.

Through a reproducibility study (Section \ref{subsec:repro}), we observe that state-of-the-art multivector methods exhibit limited efficiency. Despite their effort to optimize the gather phase, this step remains fundamentally inadequate. As emerges in recent works~\cite{emvb,warp,igp}, token-level gathering not only incurs high computational costs but also lacks the selectivity required for a compact candidate set, often returning hundreds or thousands of documents that must undergo expensive full MaxSim scoring. Driven by this observation, in contrast to the approach taken by XTR~\cite{xtr}, we take a different direction: we propose to replace the token-level gather step entirely with a \emph{document-oriented} gather strategy, where candidates are selected based on single-vector document representations.  
This design naturally aligns with the classical \emph{two-stage retrieval} architecture.
In literature, \bmtf\ and similar lexical bag-of-words models have been widely used for first-stage retrieval.
However, the limited semantic coverage of \bmtf comes at a price:
if the initial candidate set omits a large fraction of relevant documents, no reranking strategy can fully recover them. Indeed, a recent work shows that the effectiveness achieved by reranking \bmtf candidates using multivectors is limited~\cite{plaid_repro}.

\begin{figure}[t]
\centering
\includegraphics[width=\textwidth]{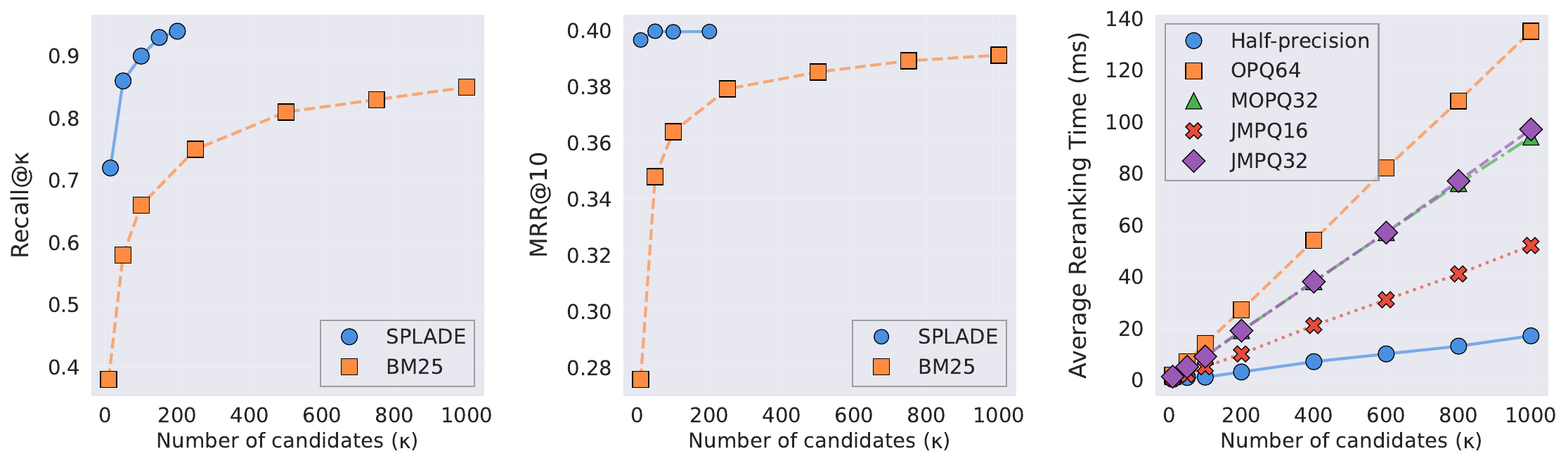}
\caption{1) Recall@$\kappa$ on \msmarcovo. 2) MRR@10 after reranking with \colbertvt on \msmarcovo. 3) Time for reranking $\kappa$ candidates for different multivectors compression schemes. \colbertvt on \msmarcovo.
} \label{fig:bm25_comparison}
\end{figure}

We claim that recent advances in first-stage retrievers based on learned sparse representations make this shift particularly effective, yielding more semantically coherent candidate sets while drastically reducing retrieval cost.
To support this, Figure~\ref{fig:bm25_comparison} (left) compares the recall at various cut-offs~$\kappa$ between \bmtf\ and a learned sparse representation obtained with the \splade\ \cocond EnsembleDistill model on the \msmarcovo\ passage dataset. The recall gap is substantial. To compensate for the recall deficit of \bmtf, pipelines typically increase the cut-off $\kappa$ to hundreds or even thousands of documents, which in turn inflates reranking cost. From Figure~\ref{fig:bm25_comparison} (right), it is clear that reranking time grows rapidly with the number of candidates, making larger cut-offs increasingly inefficient.

Even if with large $\kappa$ \bmtf\ manages to reach quite high recall, Figure~\ref{fig:bm25_comparison} (center) shows that the resulting MRR@10 after reranking is still much lower. This indicates that \bmtf\ misses many semantically relevant candidates and that increasing $\kappa$ introduces noisy documents whose scores hurt reranking quality.


In this paper, we analyse a modern two-stage pipeline where both stages use neural representations: a \emph{learned sparse retriever} for the first stage and a \emph{dense multivector late-interaction} reranker. Our contributions are:
\begin{itemize}
  \item We reproduce several state-of-the-art multivector retrieval methods on two publicly available datasets, providing a clear picture of the current multivector retrieval field.
  \item We introduce a hybrid architecture that couples LSR with multivector reranking.
  Experiments on two public datasets show that our hybrid pipeline consistently achieves one order of magnitude speedup at comparable or superior effectiveness, reaching up to $24\times$ over the best-performing competitor under similar memory and encoder settings.
  \item To enable full reproducibility of our findings, we provide a public---end-to-end---implementation of our proposed retrieval pipeline. Our implementation is available on \href{https://github.com/TusKANNy/papers_reproducibility/tree/main/ecir2026}{GitHub}.
  
\end{itemize}

\section{Related Work}
\label{sec:relwork}

\noindent \textbf{Multivector Retrieval via Late-Interaction}.
Late-interaction models such as \colbert~\cite{colbert} and \xtr~\cite{xtr} represent each document as a set of token-level vectors and compute relevance through the $MaxSim$ operator, which allows every query token to interact with every document token. This fine-grained interaction delivers state-of-the-art effectiveness but makes exhaustive retrieval computationally infeasible.

\vspace{1mm}
\noindent \textbf{Efficient Multivector Retrieval Systems}.
To address this inefficiency, early systems implemented token-level gather, which retrieves documents based on token embedding scores.
Several works have proposed mechanisms to make multivector retrieval efficient by improving either the gather or the refine step.
DESSERT~\cite{dessert} improves the gather phase using hashing-based approximate scoring, which leads to a more efficient refine phase.
\plaid~\cite{plaid} clustered token embeddings into centroids, enabling inverted-file filtering before refinement, while \colbertvt~\cite{colbertv2} introduced residual compression and improved training.
CITADEL~\cite{citadel} proposed \emph{conditional token interaction} via dynamic lexical routing, allowing each query to estimate document relevance by focusing on a subset of relevant document tokens. 
\emvb~\cite{emvb} further optimized \colbert-style retrieval through bit-vector prefiltering, SIMD acceleration, and PQ-based~\cite{pq} refinement, achieving up to $2.8\times$ faster retrieval than \plaid, with minimal quality loss.
\warp~\cite{warp} explored a complementary path, combining \colbertvt’s centroid interaction with XTR’s lightweight architecture~\cite{xtr}, achieving $3\times$ speed-ups over \colbertvt/\plaid. 
\igp~\cite{igp} introduced an incremental greedy probing strategy over proximity-graph indexes, which incrementally expands search regions and finds high-quality candidates with fewer probes.
A different line of work, MUVERA~\cite{muvera}, reformulates multivector retrieval as single-vector retrieval by partitioning the token space and aggregating embeddings into compositional vector codes, achieving high efficiency but at the cost of flexibility and fine-grained scoring.


\vspace{1mm}
\noindent \textbf{Learned Sparse Retrieval}.
Learned Sparse Retrieval (LSR) approaches bridge the gap between lexical and dense models by generating sparse term-weight representations that capture contextual semantics beyond exact token matching~\cite{sparse1,sparse2,sparse3}.  
Early works such as UniCOIL-T5~\cite{unicoil1,unicoil2} expanded document text with contextual terms predicted by DocT5Query~\cite{doc2query}, enabling lexical-style indexing with neural enrichment.  
A cornerstone in this family is \splade~\cite{splade}, which introduced sparsity-inducing regularization and log-saturation on term weights to produce highly sparse, interpretable representations while preserving semantic coverage.  
Subsequent variants refined its efficiency and regularization objectives~\cite{formal2021splade,lassance2024splade,formal2024towards}, further improving effectiveness and scalability.  
More recently, sparse retrieval has been extended beyond text to multimodal settings~\cite{multimodal_sparse}.
A promising line of investigation for LSR as first-stage retrieval is the ``Inference-Less'' paradigm, that removes the query encoder and computes query weights from fixed, precomputed lookup tables.
Early implementations adopted the simplest strategy of assigning a uniform score of 1 to all query terms~\cite{formal2021splade,lassance2024splade}.
Subsequent work introduced more sophisticated scoring mechanisms like integrating an Inverted Document Frequency (IDF) term into the loss function and use term IDF as query scores~\cite{DBLP:journals/corr/abs-2411-04403}.  Shen \emph{et al.} introduced a $l_0$ regularization to balance sparsity and avoid documents to collapse to all zero components~\cite{shen2025exploring}. 
Learned Inference-less Sparse Retrieval (\lilsr)~\cite{li-lsr} learns an optimal lookup table $\text{term} \rightarrow \text{score}$ at training time by projecting the static word embeddings into a scalar value by means of a linear layer. 

\vspace{1mm}
\noindent \textbf{Hybrid Architectures}.
Other research explicitly combines sparse and dense models. For instance, BGE-M3~\cite{bge} encoder offers support to dense, sparse and multivector retrieval seamlessly, while SparseEmbed~\cite{sparseembed} extends \splade with dense embeddings.
Approaches like SLIM~\cite{slim} and \splate put together sparse and multivector representations. \splate~\cite{splate} learns a sparse adapter on top of frozen \colbert embeddings via distillation from \splade, enabling \colbert-style reranking with sparse, interpretable candidate generation.

\vspace{1mm}
\noindent \textbf{Efficient Engines for Large-scale Retrieval}.
Modern retrieval pipelines rely on efficient Approximate Nearest Neighbors (ANN) search or inverted-index engines to support fast retrieval.
\kannolo~\cite{kannolo} is a research-oriented ANN library that supports both dense and sparse retrieval. Its sparse-HNSW implementation has shown~\cite{kannolo} to outperform the winners of the ``Sparse Track''  at NeurIPS 2023 Big ANN challenge~\cite{annneurips23}, thus representing the state-of-the-art among graph indexes for sparse retrieval.
Complementary to graph-based ANN approaches, another family of systems leverages inverted indexes optimized for LSR, including Block-Max Pruning~\cite{bmp}, Dynamic Superblock Pruning~\cite{dsbp}, and \seismic~\cite{seismic,seismic_knn,seismic_journal}. \seismic\ represents the current state-of-the-art in LSR. It partitions each inverted list into geometrically cohesive blocks represented by summary vectors. At query time, it ranks blocks via their summaries and fully evaluates only those above a relevance threshold, greatly reducing computation while maintaining accuracy.



\section{Methodology}
\label{sec:meth}

In this work, we aim to reproduce and analyze the main multivector retrieval methods proposed in the recent literature.
Our goal is to clarify their relative strengths, weaknesses, and trade-offs between efficiency and effectiveness under a unified evaluation framework.
Moreover, we extend experimental evaluation with our hybrid two-stage retrieval architecture that integrates learned sparse retrieval for candidate generation and multivector reranking for refinement.

\subsection{Two-stage Retrieval}
In our two-stage framework, we employ  \emph{Learned Sparse Retrieval} (LSR) methods as the first-stage retrievers to collect $\kappa$ candidate documents. The choice of LSR over dense single-vector representations has many motivations: 1) LSR generally offers similar performance to dense retrieval with lower memory overhead; 2) LSR can leverage inferenceless methods to remove the need for a query encoder; 3) LSR supports efficient inverted index retrieval~\cite{seismic,bmp}. The candidate set is then re-ranked using a multivector representation, from which we extract the top $\kappa_f=10$ results.

\vspace{1mm}
\noindent \textbf{Encoders}.
We adopt the \splade\ \cocond model as our sparse encoder~\cite{cocondenser} (MRR@10 on \msmarco $38.3$, Success@5 on \lotte $69.0$), and \colbertvt\ as our dense multivector encoder~\cite{colbertv2}. Both methods are reference baselines for sparse and multivector encoders.
For the inference-less sparse embeddings, we adopt the \lilsrbig version of the \lilsr model proposed in \cite{li-lsr} (MRR@10 on \msmarco $38.8$, Success@5 on \lotte $65.7$).

\vspace{1mm}
\noindent \textbf{First-Stage Retrievers.}
\kannolo and \seismic are the state-of-the-art solutions for sparse graphs and inverted index for retrieving over LSR representations~\cite{kannolo,seismic,raspberry,scalabiliy}. Hence, we pick them as our retrieval engines.
In the inference-less setting, we rely exclusively on \seismic, since, as noted in~\cite{seismic}, graph-based approaches tend to be less effective when the query is much sparser than the documents.

\vspace{1mm}
\noindent \textbf{Implementation}.
We implement the reranking in the \kannolo library by introducing the \texttt{RerankIndex} structure.  
This module exposes a primitive function $\verb|search(|q, \kappa, \kappa_f\verb|)|$ that retrieves an approximate top-$\kappa$ candidate set $\mathcal{C}$ for a query $q$ using a graph $\mathcal{G}$, and then reranks $\mathcal{C}$ via the second-stage representations.
The \texttt{RerankIndex} handles the following functionalities: 
(i) multivector data ingestion and storage, 
(ii) an optimized $MaxSim$ operator for late interaction exploiting SIMD instructions, 
(iii) quantization schemes specifically tailored for multivector embeddings, and 
(iv) a dedicated reranking index that seamlessly integrates first-stage retrieval and multivector reranking.



When employing \seismic as the first-stage retriever, we emulate the same pipeline by running a grid-search over \seismic's parameters to identify the fastest configurations achieving given accuracy thresholds.  
For each of these configurations, we rerank progressively larger candidate sets to assess the efficiency–effectiveness trade-off. We use our newly introduced \texttt{RerankIndex} to rerank the results of \seismic search.

\vspace{1mm}
\noindent \textbf{Multivector Compression}.
In our two-stage scenario, we evaluate four different compression configurations for multivector embeddings, representing a broad spectrum of memory--efficiency--effectiveness trade-offs:

\noindent \emph{Half-precision}: This setting provides maximum accuracy as it uses 16-bit floating-point per vector dimension. As \colbert embeddings have $128$ dimensions, which results in 256 bytes per token embedding, this representation suffers a high memory consumption.

\noindent \emph{OPQ}: Optimized Product Quantization~\cite{opq} is a variant of PQ~\cite{pq} that rotates the subspaces to improve compression. We use $M=64$ subspaces, resulting in $64$ bytes per token embedding. This setting reduces memory by $4\times$ but increases reranking latency due to subspace-wise distance precomputation. 

\noindent \emph{MOPQ}: Multivector-OPQ uses i) $\kappa$-Means to compute first-level centroids, and ii) OPQ with $M=32$ to compress the residuals between the original vector and the centroids. This approach takes $4$ bytes to store the centroid id, and 32 bytes for OPQ-encoded residuals, resulting in 36 bytes per token (a $7\times$ reduction).

\noindent \emph{JMPQ}: Jointly-optimized Multivector Product Quantization~\cite{jmpq}, is a supervised version of Multivector-PQ that jointly optimizes both levels and the query encoder to minimize the reconstruction error. The model we used was trained on \msmarco. We test two configurations: 16 and 32 subspaces, resulting in 20 and 36 bytes per token ($12.8\times$ and $7\times$ reductions, respectively).

The compressed configurations require an additional preprocessing phase. OPQ involves relatively fast training since it operates on small subspaces, whereas both MOPQ and JMPQ employ $\kappa$-means clustering at the first level, introducing a notable preprocessing overhead. JMPQ further adds complexity, as it requires end-to-end joint optimisation during training.



\vspace{1mm}
\noindent \textbf{Reranking Optimizations}.
We propose two novel lightweight optimization strategies to improve reranking efficiency.
\noindent \emph{Candidates Pruning (CP)}: we truncate the candidate list using a threshold $\alpha$ based on the difference between the $\kappa_f$-th candidate score and subsequent scores. A sharp drop in first-stage scores suggests a general relevance decline, allowing us to safely discard lower-ranked documents. More precisely, let $t$ be the first-stage score of the $\kappa_f$th candidate; if, for the first-stage score $s$ of a candidate it holds $s < (1-\alpha)t$, then we discard the current candidate and all the other candidates below it.
\noindent \emph{Early Exit (EE)}: during reranking, if the top-$\kappa_f$ results remain unchanged for $\beta$ consecutive candidates, we stop reranking early, speculating on the fact that successive candidates are unlikely to improve the final ranking. In particular, if for $\beta$ steps the set of $\kappa_f$ best results so far is not updated after computing late interaction score, then stop reranking and return the current set of best results.

\vspace{1mm}
\noindent \textbf{Parameters Selection}.
When building HNSW with \kannolo, we experimentally verify that number of neighbors $M=32$ and construction precision $ef_c \in \{1500, 2000\}$ yield optimal results respectively on \lotte and \msmarco.
At search time, the \emph{search expansion factor} ($ef_s$) tunes the efficiency-effectiveness tradeoff. Thus, we perform a complete grid search over the set $\{ef_s, \kappa, \alpha, \beta \}$. 

For \seismic, we first tune its construction and search parameters to reach Accuracy@50 (the fraction of true nearest neighbors contained within the top-50 approximate retrieval results) levels in $[0.90, 0.99]$ with $0.01$ steps. For each configuration, we then perform offline reranking at different cut-offs of $\kappa \leq 50$ and sum the first- and second-stage latencies.

The reranking hyperparameters tested are: $\kappa \in \{ 15, 20, 25, 30, 35, 40, 50 \}$, $\alpha \in \{ 0.015, 0.025, 0.050, \text{\paroff} \}$, $\beta \in \{ 2, 3, 4, \text{\paroff} \}$, where \lq\lq \paroff\rq\rq \ means that the optimization (CP or EE) is disabled.

\subsection{Competitors}
We report the detailed experimental configuration of the evaluated methods. Whenever possible, we rely on the pre-built indexes releases by the authors, to ensure a fair reproduction of their results and avoid crafting ourselves sub-optimal configurations.


\vspace{1mm}
\noindent \textbf{EMVB.} Indexes are publicly available on the official GitHub repository.\footnote{\url{https://github.com/CosimoRulli/emvb}}
While the authors provide the hyperparameters to replicate the original results, we perform a grid search to validate this approach further. The specific ranges for each parameter can be found in our code. EMVB is written in C++. 

\vspace{1mm}
\noindent \textbf{IGP.} Indexes are also publicly available on their GitHub repo.\footnote{\url{https://github.com/DBGroup-SUSTech/multi-vector-retrieval}}
The authors in the paper indicate specific ranges for both search parameters. In particular, they recommend $\phi_{pb} \in \{1, 2, 4, 8, 16, 32, 64\}$ and $\phi_{ref} \in \{1, 2, 5, 10, 20, 30, 40, 50 \} \cdot k_f$. IGP is written in C++. 


\vspace{1mm}
\noindent \textbf{WARP.}
\warp's authors did not test their method on \msmarco and they employ \lotte only on XTR's encodings, so using \warp's code we created our own encodings and built our own indexes. 
For the construction parameters, the authors suggest $b\in\{2,4\}$ bits per component to encode the residual embeddings. We choose $b=2$ in order for us to have a direct comparison against \igp (2 bits), \emvb (MOPQ32, resulting in 2 bits per component) and our MOPQ32/JMPQ32 settings. 
Other construction parameters, like the number of $\kappa$-means centroids, are empirically computed based on the characteristics of the dataset. 
Two parameters control the search quality in \warp: the number of clusters to decompress per query token $n_{probe}$ and the threshold on the cluster size used for $\text{WARP}_{\text{SELECT}}$ $t^{\prime}$. From their experiment, they find that \warp is robust to variations of $t^{\prime}$ and they choose to compute it empirically basing on dataset's size. Thus, for our grid search, we let $n_{probe}$ vary in the range $\{1,2,4,8,16,32,64 \}$ as they suggest.

\section{Experiments}
\label{sec:exp}

We conduct experiments on the \msmarco passages dataset~\cite{msmarco} for the in-domain
evaluation and on \lotte-pooled~\cite{colbertv2} for the out-of-domain evaluation. The \msmarco dataset is a collection of $8.8$\,M passages; we use the $6{,}980$ queries in the \emph{dev.small} to assess the effectiveness of our approaches, measured in terms of MRR@10. \lotte-pooled contains approximately $2.4$\,M passages drawn from multiple domains (e.g., search, forum, Wikipedia, news, and StackExchange). We use the \texttt{search/dev} split comprising $2{,}931$ queries.
We generate the multivector embeddings for both collections with the original \colbert code, using the \colbertvt checkpoint.\footnote{\url{https://github.com/stanford-futuredata/ColBERT}} Optimized Product Quantization is trained using the \faiss library~\cite{douze2024faiss}, while for JMPQ we use the original code~\cite{jmpq}.

\vspace{1mm}
\noindent \textbf{Hardware Details}. Our experiments were executed on a server equipped with 512 GiB of RAM and 2 Intel Xeon Silver 4314 clocked at $2$.$4$GHz, with $64$ total threads between physical and hyper-threaded ones.
Index construction is performed using multi-threading to exploit available CPU resources fully. In contrast, all search operations are executed sequentially
on a single core to ensure fair and consistent performance comparisons across
different indexing methods and settings. 
Our methods are implemented in Rust (cargo 1.92.0-nightly) and compiled in release mode 
to enable CPU-specific optimizations.

\subsection{Reproducibility and Comparison with the State of the Art}\label{subsec:repro}
Tables~\ref{tab:comp_msmarco} and~\ref{tab:comp_lotte} compare our reranking methods against \emvb, \warp, and \igp\ in terms of memory usage (bytes per token vector), effectiveness (MRR@10 or Success@5), and retrieval time (ms) on \msmarcovo\ and \lotte-pooled, respectively.  

For our first stage, we consider three variants: i) Double-encoder - \kannolo, where we employ the end-to-end RerankIndex implemented in \kannolo and the first stage relies on \splade \cocond representations; ii) Double-encoder - \seismic, where the first stage still relies on \splade \cocond as a sparse encoder, but we use \seismic for first-stage retrieval; iii) \lilsr \ - \seismic, that replaces \splade \cocond with the \lilsr \lilsrbig model. The reranking is tested with the half-precision baseline, JMPQ16/32 (on \msmarco) and MOP32 (on \lotte).

\begingroup
\setlength{\tabcolsep}{7pt} 
\begin{table}[t]
\centering
\caption{Retrieval time (ms) on \msmarcovo ($-$ indicates not reached). Bytes per vector for \warp\ are estimated from index size.
\label{tab:comp_msmarco}}
\begin{tabular}{
    l
    @{\hspace{1.2em}}
    S[table-format=3.0,table-number-alignment=right]
    @{\hspace{0.8em}}
    *{6}{S[table-format=3.1,table-number-alignment=right]}
}
\toprule
& \textbf{\multirow{2}{*}{Bytes}}
& \multicolumn{6}{c}{\textbf{MRR@10}}\\
\cmidrule(lr){3-8}
& & \multicolumn{1}{c}{0.390} & \multicolumn{1}{c}{0.392}
  & \multicolumn{1}{c}{0.394} & \multicolumn{1}{c}{0.396}
  & \multicolumn{1}{c}{0.398} & \multicolumn{1}{c}{0.399} \\
\midrule
\warp       & 35  & 98.7 & 98.7 & 136.3 & \na & \na & \na \\
\igp              & 36 & 72.8 & 110.0 & \na & \na & \na & \na \\
\emvb (JMPQ16)    & 20  & 56.0 & 56.7 & 61.4 & \na & \na & \na \\
\emvb (JMPQ32)    & 36  & 56.0 & 56.0 & 56.4 & 57.3 & 59.4 & 67.1 \\
\midrule
\multicolumn{8}{l}{Double-encoder - \kannolo}\\
\quad w/ Half-precision & 256   & 2.3 & 2.8 & 3.0 & 4.1 & 6.2 & 8.7 \\
\quad w/ JMPQ16         & 20   & 2.4 & 2.9 & 4.2 & 6.6 & \na & \na \\
\quad w/ JMPQ32         & 36   & 2.8 & 3.0 & 3.4 & 4.2 & 5.5 & 8.1 \\
\midrule
\multicolumn{8}{l}{Double-encoder - \seismic}\\
\quad w/ Half-precision & 256   & 0.5 & 0.5 & 0.5 & 0.7 & 1.5 & \na \\
\quad w/ JMPQ16         & 20   & 1.0 & 1.0 & 1.3 & \na & \na & \na \\
\quad w/ JMPQ32         & 36   & 1.5 & 1.5 & 1.5 & 1.5 & 2.4 & 2.5 \\
\midrule
\multicolumn{8}{l}{\lilsr \  - \seismic}\\
\quad w/ Half-precision & 256   & 1.5 & 1.5 & 1.7 & 2.3 & 5.4 & \na \\
\quad w/ JMPQ16         & 20   & 2.3 & 2.9 & 5.8 & \na & \na & \na \\
\quad w/ JMPQ32         & 36   & 2.4 & 2.4 & 2.4 & 2.6 & 5.7 & \na \\
\bottomrule
\end{tabular}
\end{table}
\endgroup

\begingroup
\setlength{\tabcolsep}{7pt} 
\begin{table}[!h]
\centering
\caption{Retrieval time (ms) on \lotte-pooled ($-$ indicates not reached). Bytes per vector for \warp\ are estimated from index size.
\label{tab:comp_lotte}}
\begin{tabular}{l@{\hspace{1.2em}} S[table-format=3.0] @{\hspace{0.8em}}
                *{6}{S[table-format=2.1,table-number-alignment=right]}}
\toprule
& \textbf{Bytes} & \multicolumn{6}{c}{\textbf{Success@5}}\\
\cmidrule(lr){3-8}
& & \multicolumn{1}{S}{67.0} & \multicolumn{1}{S}{67.5}
    & \multicolumn{1}{S}{68.0} & \multicolumn{1}{S}{68.5}
    & \multicolumn{1}{S}{69.0} & \multicolumn{1}{S}{69.5} \\
\midrule
\warp       & 35  & 49.9 & 49.9 & \na & \na & \na & \na \\
\igp              & 36  & 39.0 & 48.7 & \na & \na & \na & \na \\
\emvb (MOPQ32)    & 36  & 54.0 & 54.7 & 55.4 & 59.2 & 74.2 & \na \\
\midrule
\multicolumn{8}{l}{Double-encoder - \kannolo}\\
\quad w/ Half-precision      & 256  & 1.7 & 1.8 & 2.1 & 2.4 & 3.0 & 4.0 \\
\quad w/ MOPQ32     & 36   & 4.0 & 4.0 & 4.3 & 4.7 & 5.8 & 10.1 \\
\midrule
\multicolumn{8}{l}{Double-encoder - \seismic}\\
\quad w/ Half-precision      & 256  & 1.3 & 1.3 & 1.3 & 1.3 & 1.3 & 1.4 \\
\quad w/ MOPQ32     & 36   & 3.1 & 3.1 & 3.1 & 3.1 & 3.1 & \na \\
\midrule
\multicolumn{8}{l}{\lilsr \ - \seismic}\\
\quad w/ Half-precision      & 256  & 3.3 & 3.3 & 3.3 & 3.3 & 3.7 & \na \\
\quad w/ MOPQ32     & 36   & 5.0 & 5.0 & 5.0 & 5.4 & \na & \na \\
\bottomrule
\end{tabular}
\end{table}
\endgroup

In both tables, we report the cost of storing a term vector for all the multi-vector methods (\emvb, \igp, \warp). For our reranking methods, observe that storing the sparse embeddings induces additional memory overhead. On \msmarco, the \splade \cocond embeddings have a size of approximately 4 GB, whereas the inferenceless variant requires about 13 GB. On \lotte, the corresponding sizes are approximately 1.5 GB and 4.2 GB for \splade \cocond and inferenceless data, respectively. For completeness, the size of the multivector datasets is 22.5 GB and 13.5 GB, respectively for JMPQ32 and JMPQ16 on \msmarco, and 10 GB for MOPQ32 on \lotte. The storage required by the index structures themselves is negligible and therefore omitted for both sparse and multivector indexes.



\vspace{1mm}
\noindent \textbf{Comparative Results Analysis}.
On \msmarco (Table~\ref{tab:comp_msmarco}), \emvb offers superior performance than both \igp and \warp, with both lower latencies and higher peak effectiveness. On \lotte(Table~\ref{tab:comp_lotte}),  \emvb has slightly higher latencies than \warp and \igp but still yields higher peak effectiveness.
Overall, our reproduced performance aligns well with the results reported in the original \emvb paper.
\warp, though originally designed for \xtr, performs comparably to \igp, confirming the generalization capability to \colbertvt\ that was noted by its authors in the appendix of~\cite{warp}.  

On \msmarco, \igp\ demonstrates effectiveness levels consistent with those reported in the original paper~\cite{igp}, although its reported latencies are significantly lower than those reported for \emvb; in our experiments, however, we observe the opposite trend, with \emvb consistently outperforming \igp. The absolute differences between our measured times and those reported in the original paper may be due to different hardware configurations, while the reversed trend likely arises from the authors not using the optimized indexes provided in \emvb's official repository nor its original index-building code.\footnote{The authors implemented their own index building \url{https://tinyurl.com/3yd8fjam}.}
We highlight that in the \igp original paper, the authors evaluated \igp on different metrics, specifically on MRR and Recall, while we attain to use the \lotte official metric~\cite{colbertv2}, i.e., Success@5.

Across both datasets, all three of our configurations are approximately one order of magnitude faster than the competitor methods at comparable memory budgets. This advantage is due to the small set of candidates we need to rerank ($\leq 50$ in all our settings) and the efficient first-stage retrievers we employed. 
Because our reranking stage operates on a small candidate set—and our optimizations (CP, EE) further reduce the number of full evaluations—the pipeline’s bottleneck shifts to the first-stage retrieval. Consequently, in the double-encoder setting, \seismic\ provides significant benefits over \kannolo, especially at higher effectiveness levels where latency becomes dominated by the first stage.
The inferenceless configuration exhibits higher latency due to heavier vectors and the larger number of non-zero components, which increase index size and slow similarity computations. However, this setup is directly comparable to the competitors since it uses only one encoder at query time.  
Under this fair comparison (i.e., with only one encoder involved), the inferenceless compressed setting achieves speed-ups of $10\times$–$24\times$ on \msmarco\ and $7.8\times$–$11\times$ on \lotte\ relative to the fastest competing method.

\vspace{1mm}
\noindent \textbf{Reproducibility Remarks}.
Overall, our reproduced results are often consistent with the original findings.
\emvb, though preceding \igp\ and \warp, shows stable and competitive performance in both latency and effectiveness, confirming the authors’ reported results.
\warp\ generalises beyond its native \xtr\ setup. However, a direct comparison to \emvb was missing in prior work, and we close this gap with our evaluation.
\igp\ reproduces the reported effectiveness on \msmarco\ but exhibits higher latencies relative to \emvb, likely due to the use of different indexes and modified build procedures. Moreover, its evaluation does not include \lotte's official metric (Success@5).

\subsection{Impact of Quantization and Reranking Optimizations}

\begin{figure}[!t]
\centering
\includegraphics[width=\textwidth]{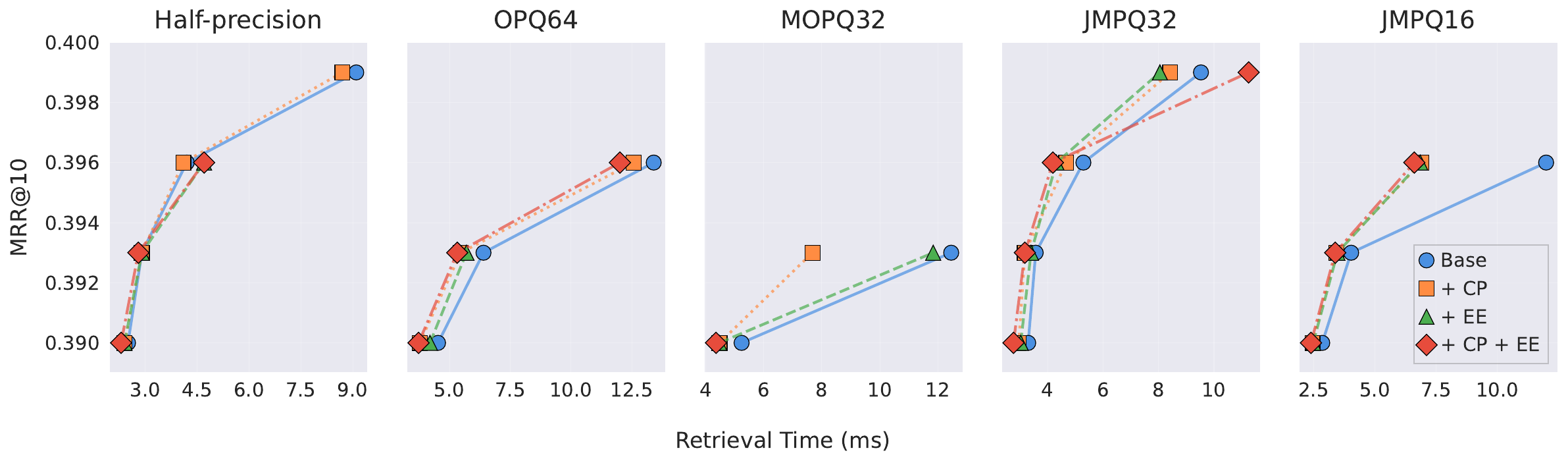}
\caption{Evaluation of quantization schemes and optimizations. End-to-end (1st + 2nd stage) retrieval time (ms) on \msmarco.\vspace{-5mm}} \label{fig:ablation_quant}
\end{figure}

We perform an ablation study on the \msmarco\ dataset to analyze the effect of multivector quantization and our proposed reranking optimizations, i.e. \emph{Candidate Pruning} (CP) and \emph{Early Exit} (EE).  
Figure~\ref{fig:ablation_quant} reports MRR@10 versus search latency for all tested configurations on the \msmarco dataset, where on first stage we employ \splade \cocond representations.

The figure compare all five compression schemes, i.e., Half-precision, OPQ64, MOPQ32, JMPQ32 and JMPQ16, under different optimization settings.  
JMPQ32 shows no loss in quality despite a $7\times$ memory reduction compared to the Half-precision setting. The other compressed variants exhibit slightly lower MRR but deliver substantial memory savings, ranging from $4\times$ (OPQ64) to $12.8\times$ (JMPQ16) relative to the baseline. 
OPQ64 maintains strong accuracy but suffers from the highest latency (up to 13.5\,ms) due to the computation of 64 distance tables per query token.  
MOPQ32 achieves similar performance at roughly half the memory footprint (36 Bytes per embedding), although its peak effectiveness is slightly lower.
JMPQ32 and JMPQ16 deliver the best overall trade-offs: they retain most of the effectiveness while improving efficiency and compressing memory by $7\times$--$12.8\times$.

Regarding the reranking optimizations, CP and EE consistently improve efficiency across all quantization settings, achieving a maximum speed-up of $1.8\times$. CP is the most reliable optimization; by relying on uncompressed first-stage scores, it consistently reaches the same MRR as its unoptimized counterpart. Conversely, EE is occasionally too aggressive, preventing the algorithm from reaching top accuracy; this also affects the combined EE + CP setting.

Overall, these results confirm that:
(i) quantization can substantially reduce memory with limited effectiveness loss,  
(ii) CP and EE provide substantial efficiency gains at no cost in effectiveness, and  
(iii) JMPQ achieves the best overall efficiency–effectiveness balance, but it needs supervised training; MOPQ and OPQ still offer valid alternatives.

\section{Conclusion}
\label{sec:conclusion}

We revisited recent multivector retrieval methods within a unified, fully reproducible framework, reproducing and comparing three state-of-the-art systems---\emvb, \warp, and \igp---under identical hardware and evaluation settings. Our reproduction confirms that, despite substantial engineering optimizations, these token-level Gather \& Refine pipelines exhibit high query latency in practice.

Although we do not decompose the internal costs of the gather and refine phases, this behavior is consistent with the design of token-level retrieval, which relies on expensive token-level indexes and retrieves hundreds or thousands of candidates for full-scoring, as documented in prior work.

Motivated by this observation, we explored an alternative design in which the gather phase operates at the document level using learned sparse retrieval, and multivector scoring is applied only as a reranking step. Our benchmarks show that this two-stage pipeline requires only $20$–$50$ candidates for reranking to match or exceed the effectiveness of gather-based multivector systems, while substantially reducing query latency. Under comparable memory and encoder settings, this approach achieves speed-ups of up to $24\times$ over the fastest reproduced baselines. Our experimental analysis further investigates quantization and novel pruning and early-exit strategies, showing that quantization effectively reduces memory usage and that adaptive reranking consistently improve efficiency.


\section{Acknoledgments}
This work was partially supported by the European Union’s Horizon 2020 Research and Innovation Staff Exchange programme under the Marie Skłodowska-Curie grant agreement No. 872539,  by the Horizon Europe RIA ``Extreme Food Risk Analytics'' (EFRA), grant agreement No. 101093026, by the PNRR - M4C2 - Investimento 1.3, Partenariato Esteso PE00000013 - ``FAIR - Future Artificial Intelligence Research'' - Spoke 1 ``Human-centered AI'' funded by the European Commission under the NextGeneration EU program, and by the MUR-PRIN 2022 ``Algorithmic Problems and Machine Learning'', grant agreement n. 20229BCXNW.

\begin{credits}
\subsubsection{\discintname}
The authors have no competing interests to declare that are relevant to the content of this article.
\end{credits}


%
%
%
%
\bibliographystyle{splncs04}
\bibliography{references.bib}

\end{document}